\documentclass[prl,twocolumn,preprintnumbers,amsmath,amssymb]{revtex4}

\usepackage[normalem]{ulem}
\usepackage{times}
\usepackage{natbib}
\usepackage{amssymb,amsbsy,amsmath,amsfonts}
\usepackage{graphicx}% Include figure files
\usepackage{epsf,epsfig,float,latexsym,amsthm,fancyhdr,rotating}
\usepackage{graphics,psfrag,longtable}
\usepackage[svgnames]{xcolor}
\usepackage{ulem}
\usepackage{slashed,wasysym}

\newcommand{\be}{\begin{equation}}
\newcommand{\ee}{\end{equation}}
\newcommand{\beq}{\begin{equation}}
\newcommand{\eeq}{\end{equation}}
\newcommand{\bea}{\begin{eqnarray}}
\newcommand{\eea}{\end{eqnarray}}
\newcommand{\ba}{\begin{eqnarray}}
\newcommand{\ea}{\end{eqnarray}}

\newcommand{\eS}{\epsilon_S}

\newcommand{\eT}{\epsilon_T}
\newcommand{\ket}[1]{| #1 \rangle}
\newcommand{\bra}[1]{\langle #1 |}
\begin{document}
\preprint{UCSD/PTH 14-11}
\title{Non-standard semileptonic hyperon decays}

\author{Hsi-Ming Chang$^{1}$}
\author{Martin Gonz\'alez-Alonso$^{2}$}
\author{Jorge Martin Camalich$^{1,3}$}

\affiliation{
$^1$Dept. Physics, University of California, San Diego, 9500 Gilman Drive, 
La Jolla, CA 92093-0319, USA\\
$^2$
IPNL, CNRS, IN2P3, 4 rue E. Fermi, 69622 Villeurbanne cedex, France; Universit\'e Lyon 1, Villeurbanne
Universit\'e de Lyon, F-69622, Lyon, France\\
$^3$PRISMA Cluster of Excellence Institut f\"ur Kernphysik, 
Johannes Gutenberg-Universit\"at Mainz, 55128 Mainz, Germany}

\begin{abstract}
We investigate the discovery potential of semileptonic hyperon decays in terms 
of searches of new physics at teraelectronvolt scales. These decays are controlled by
a small $SU(3)$-flavor breaking parameter that allows for systematic expansions and 
accurate predictions in terms of a reduced dependence on hadronic form factors. 
We find that muonic modes are very sensitive to non-standard scalar and tensor 
contributions and demonstrate that these could provide a powerful synergy with direct
searches of new physics at the LHC. 

\end{abstract}
\maketitle

\paragraph{Introduction.-} 

The meson and baryon semileptonic decays have played a crucial role in the discovery of
the $V - A$ structure~\cite{Weinberg:2009zz} and quark-flavor mixing~\cite{Cabibbo:1963yz}
of the (charged current) electroweak interactions in the Standard Model (SM). 
From a modern perspective, high-precision measurements of these decays provide 
a benchmark to test the SM and complement the direct searches of new physics (NP) at 
teraelectronvolt (TeV) energies.

For example, the accurate determination of the elements $V_{ud}$ and $V_{us}$ of the 
Cabibbo-Kobayashi-Maskawa (CKM) matrix can be used to test its unitarity, constraining 
NP with characteristic scales as high as $\Lambda\sim10$~TeV~\cite{Cirigliano:2009wk}.
Furthermore, one can test the $V - A$ structure of the charged currents in $d\!\to\! u$
transitions using neutron and nuclear $\beta$ decays~\cite{Herczeg:2001vk,Severijns:2006dr,
Cirigliano:2009wk,Bhattacharya:2011qm,Cirigliano:2013xha,Gonzalez-Alonso:2013uqa,Gonzalez-Alonso:2013ura} and pion
decays~\cite{Herczeg:1994ur,Campbell:2003ir}. 
Current limits for the associated NP scale are also at the TeV level, and important
improvements are expected from future experiments~\cite{Cirigliano:2012ab}. 
Searches of non-standard $d\to u$ transitions can also be done using LHC data,
through e.g. the collision of $d$ and $u$ partons in the $pp\to e^{\pm}+MET+X$ channel (where MET stands for
missing transverse energy)~\cite{Cirigliano:2012ab}. This leads to an interesting synergy between low- and 
high-energy NP searches in these flavor-changing processes.

A similar comprehensive analysis of exotic effects in $s\to u$ transitions has not 
been done yet. The (semi)leptonic kaon decays are optimal laboratories for this
study due to the intense program of high-precision measurements and accurate calculations of the 
relevant form factors that has been carried out over the last 
decades~\cite{Cirigliano:2011ny}. Indeed, bounds on right-handed~\cite{Bernard:2006gy,Bernard:2007cf} or scalar and 
tensor~\cite{Yushchenko:2004zs} NP interactions at the $10^{-2}-10^{-3}$ level 
(relative to the SM) can be obtained~\cite{Antonelli:2009ws,Carpentier:2010ue}. 
Generally speaking, (pseudo)scalar and tensor operators modify the spectrum of the decay and a detailed
knowledge of the $q^2$ dependence of the form factors becomes necessary~\cite{Antonelli:2008jg}.

In this letter we investigate the physics potential of the semileptonic 
hyperon decays (SHD) to search for NP.   
Although the description of these modes may seem involved due 
to the presence of six nonperturbative matrix elements
or form factors, they present interesting 
features~\cite{Gaillard:1984ny,Garcia:1985xz,Cabibbo:2003cu,Kadeer:2005aq}:  
$(i)$~In the isospin limit, there are a total of 8 different channels, 
each having a differential decay rate with a rich angular distribution 
that could involve the polarizations of the baryons. $(ii)$~The same form
factors in different channels can be connected to each other and with
other observables (e.g. electromagnetic form factors) in a model-independent
fashion using the approximate $SU(3)$-flavor symmetry of QCD. $(iii)$~The maximal 
momentum transfer is small compared to the baryon masses and it is parametrically 
controlled by the breaking of this symmetry. Therefore, a simultaneous $SU(3)$-breaking
and ``recoil'' expansion can be performed that simplifies, systematically, 
the dependence of the decay rate on the form factors.

On the experimental side there is much room for improvement.
Except for the measurements performed by the KTeV and NA-48 Collaborations in the $\Xi^0\!\to\Sigma^+$ 
channel~\cite{Affolder:1999pe,AlaviHarati:2001xk,AlaviHarati:2005ev,Batley:2006fc,Batley:2012mi}, 
most of the SHD data is more than 30 years old~\cite{Agashe:2014kda}. 
 On the other hand, (polarized) hyperons could be produced
abundantly in the NA62 experiment at CERN~\cite{PicciniNA62} or in any other 
hadron collider like the future $p\bar{p}$ facility PANDA~\cite{Schonning:2014kza}
at FAIR/GSI or J-PARC~\cite{Miwa:2012zz}. 

In the following, we investigate the physics reach of the SHD with a discussion based on
the sensitivity of the total decay rates to non-standard scalar and tensor interactions. 
We show that the bounds from SHD are competitive with
those derived from the LHC data on the $pp\to e^{\pm}+MET+X$ channel and leave the interplay with 
kaon decays for future work (see~\cite{Bernard:2006gy,Bernard:2007cf,Yushchenko:2004zs,Antonelli:2009ws} 
for the current status).
 
\paragraph{The SM effective field theory.-}

In the SM, and at energies much lower than the electroweak symmetry breaking scale, 
$v=(\sqrt{2}G_F)^{-1/2}\simeq246$ GeV, all charged-current weak
processes involving up and strange quarks are described by the Fermi $(V-A)\times(V-A)$ four-fermion interaction. 
Beyond the SM, the most general effective Lagrangian is~\cite{Cirigliano:2009wk}:
\begin{eqnarray}
{\cal L}_{\rm eff} 
&=&
- \frac{G_F V_{us}}{\sqrt{2}} \,\Big(1 + \epsilon_L + \epsilon_R  \Big)\times\nonumber\\
&&\sum_{\ell=e,\mu}\Big[
\bar{\ell}  \gamma_\mu  (1 - \gamma_5)   \nu_{\ell} 
\cdot \bar{u}   \Big[ \gamma^\mu \ - \  \big(1 -2  \epsilon_R  \big)  \gamma^\mu \gamma_5 \Big] s \nonumber\\
&+& \bar{\ell}  (1 - \gamma_5) \nu_{\ell}
\cdot \bar{u}  \Big[  \epsilon_S  -   \epsilon_P \gamma_5 \Big]  s\nonumber\\
&+& \epsilon_T     \,   \bar{\ell}   \sigma_{\mu \nu} (1 - \gamma_5) \nu_{\ell}    \cdot  \bar{u}   \sigma^{\mu \nu} (1 - \gamma_5) s
\Big]+{\rm h.c.}, \ \ \ \  \
\label{eq:leffq2} 
\end{eqnarray}
neglecting ${\cal O}(\epsilon^2)$ terms and derivative interactions, and where we
use $\sigma^{\mu \nu} =  [\gamma^\mu, \gamma^\nu]/2$. This Lagrangian has 
been constructed using only the SM fields relevant at low scales, $\mu\sim1$ GeV,
and demanding the operators to be color and electromagnetic singlets. Furthermore,
we have restricted our attention to non-standard interations that conserve lepton flavor 
and are lepton universal. Finally, we assume that the Wilson coefficients (WC) $\epsilon_i$ are real, since we focus on $CP$-even observables.  

In light of the  null results in direct searches of NP at colliders, we assume that its typical
scale, $\Lambda$, is much larger than $v$. In such case, NP can be parameterized using an 
effective (non-renormalizable) Lagrangian, ${\cal L}_{\rm eff} = {\cal L}_{\rm SM} + (1/\Lambda^2) \sum_i \alpha_i {\cal O}_i^{(6)} + \ldots ~,$
where the ${\cal O}_i^{(6)}$ are now operators built with \textit{all} the SM fields and subject to the structures
of its full (unbroken) gauge symmetry group~\cite{Buchmuller:1985jz}. 
The WC $\epsilon_i$ in eq.~(\ref{eq:leffq2}) are generated by the high-energy WC $\alpha_i$, 
which in turn can be obtained by matching to a particular NP model at $\mu=\Lambda$, and by running 
down to $\mu\sim1$ GeV using the renormalization group equations, with the heavier fermions
and weak bosons integrated out in the process~\cite{Grzadkowski:2010es,Jenkins:2013zja,Jenkins:2013wua,Alonso:2013hga,Alonso:2014zka}.

This framework, usually referred to as the SM effective field theory (SMEFT), allows for a
bottom-up investigation of NP, describing the implications of collider searches for low-energy 
experiments and vice versa. Needless to say, this interplay would become crucial in shaping
the NP if a discrepancy with the SM is to be found. Examples of top-down applications, with 
correlated effects at high- and low-energies, can be found in scenarios with 
lepto-quarks~\cite{Davidson:2010uu} or extra scalar fields~\cite{Antonelli:2008jg,Bhattacharya:2011qm}.

\paragraph{Semileptonic hyperon decays.-} Neglecting electromagnetic corrections,
the amplitude for a particular SHD $B_1 (p_1) \to B_2 (p_2) \ell^- (p_\ell) \bar{\nu}_\ell (p_\nu)$
factorizes into the leptonic and baryonic matrix elements. For the (axial)vector hadronic currents we 
have the parametrization in terms of the standard form factors~\cite{Cabibbo:2003cu,Weinberg:1958ut}:
 \begin{eqnarray}
 &&\bra{B_2 (p_2) } \bar{u} \gamma_\mu s \ket{B_1 (p_1)} 
 =
 \bar{u}_2 (p_2)  \Big[
 f_1(q^2)  \,  \gamma_\mu    
 \nonumber\\
 &&+ \frac{f_2(q^2)}{M_1}   \, \sigma_{\mu \nu}   q^\nu  
 + \frac{f_3(q^2)}{M_1}   \,  q_\mu  
 \Big]  
  u_1 (p_1),  \label{eq:vectorFF}\\
 &&\bra{B_2 (p_2) } \bar{u} \gamma_\mu \gamma_5  s \ket{B_1 (p_1)} 
 =
 \bar{u}_2 (p_2)  \Big[
 g_1(q^2)    \gamma_\mu    \nonumber\\
 &+& \frac{g_{2} (q^2)}{M_1}   \sigma_{\mu \nu}   q^\nu  
 + 
 \frac{g_{3} (q^2)}{M_1}   q_\mu  
 \Big]   \gamma_5  u_1 (p_1),  \label{eq:axialvectorFF}
 \end{eqnarray}
whereas the non-standard (pseudo)scalar and tensor interactions introduce
new form factors~\cite{Weinberg:1958ut}:
\begin{eqnarray}
&&\hspace{-1cm}\bra{B_2 (p_2) } \bar{u} \,   s \ket{B_1 (p_1)} 
=
f_S(q^2)  \ \bar{u}_2 (p_2)  \, u_1 (p_1), \label{eq:scalarFF} \\ 
&&\hspace{-1cm}\bra{B_2 (p_2) } \bar{u} \,  \gamma_5 \,  s \ket{B_1 (p_1)}=
g_P(q^2)  \ \bar{u}_2 (p_2)  \, \gamma_5 \, u_1 (p_1),  \label{eq:pseudoscalarFF} \\
&&\hspace{-1cm}\bra{B_2 (p_2) } \bar{u} \, \sigma_{\mu \nu}  \,  s \ket{B_1 (p_1)} 
\simeq
f_T(q^2) \,  \bar{u}_2 (p_2)\,
\sigma_{\mu \nu}\,u_1 (p_1). \label{eq:tensorFF}  
\end{eqnarray}
In eqs.~(\ref{eq:vectorFF})-(\ref{eq:tensorFF}), $u_{1,2}$ are the parent and
daughter baryon spinor amplitudes, $M_{1,2}$ their respective masses, $q = p_1 - p_2$
is the momentum transfer,  with $m_\ell^2\leq q^2\leq(M_1-M_2)^2$. Furthermore, in Eq. (\ref{eq:tensorFF}) we 
have neglected other contributions to the matrix element of the tensor current since they 
are kinematically suppressed $\sim\mathcal{O}(q/M_1)$~\cite{Weinberg:1958ut}. 

A crucial aspect in the study of the SHD is the approximate $SU(3)$-flavor symmetry
of QCD. It controls the phase space of the decay and allows for a systematic expansion
of the observables in the \textit{generic} symmetry breaking parameter, 
$\delta=(M_1-M_2)/M_1$~\cite{Garcia:1985xz}. 
Relations among form factors are obtained in the exact symmetric limit using standard
group theory~(see e.g. Ref.~\cite{Georgi:1982jb}) and $\mathcal{O}(\delta)$ corrections can be calculated using
model independent methods~\cite{Krause:1990xc,Jenkins:1990jv,Geng:2008mf,Ledwig:2014rfa,Geng:2013xn,Guadagnoli:2006gj,Sasaki:2008ha,Sasaki:2012ne}.
In addition, the form factors can be expanded around $q^2=0$ in powers of $q^2/M_X^2\sim\delta^2$,
where $M_X\sim1$ GeV is a hadronic scale related to the mass of the resonances coupling to
the currents~\cite{Ecker:1989yg,Masjuan:2012sk}.

Let us illustrate this with the total decay rate for the
electronic mode in the SM which, expanded up to next-to-leading order (NLO) in $\delta$ and
neglecting $m_e$, is~\cite{Garcia:1985xz}:
\begin{eqnarray}
&&\Gamma_{e,\,{\rm SM}}\simeq\frac{G_F^2\,|V_{us}\,f_1(0)|^2\,\Delta^5}{60\,\pi^3}\,\left[\left(1-\frac{3}{2}\delta\right)\nonumber \right.\\
&&\left.+3\,\left(1-\frac{3}{2}\delta\right)\frac{g_1(0)^2}{f_1(0)^2}-
4\delta\frac{g_2(0)}{f_1(0)}\frac{g_1(0)}{f_1(0)}\right],\label{eq:Gammae}
\end{eqnarray}
with $\Delta=M_1-M_2$. This expression contains a minimal dependence on the form factors.
No information on their $q^2$ dependence is required and, moreover, the last term can be
neglected because the weak-electric charge, $g_2(0)$, is itself $\mathcal{O}(\delta)$~\cite{Weinberg:1958ut}.
Thus, besides $G_F$ and $V_{us}$, and up to a theoretical accuracy of $\mathcal{O}(\delta^2)\sim 1-5\%$, 
the total decay rate of the electronic mode in the SM only depends on hyperon vector and axial
charges, $f_1(0)$ and $g_1(0)$. Eq.~(\ref{eq:Gammae}) makes manifest that
$f_1(0)$ is essential for extracting $V_{us}$ from the rates, while the ratio
$g_1(0)/f_1(0)$ can be obtained measuring the angular distribution of the final
lepton~\cite{Garcia:1985xz,Cabibbo:2003cu}. Neglected electromagnetic
corrections are of a few percent~\cite{Garcia:1985xz,FloresMendieta:1996np}, well within the accuracy
achieved at NLO in the $SU(3)$ expansion.

Beyond the SM, we generally have two types of effects. On one hand, (axial)vector
modifications to the SM, described by the WC $\epsilon_{L,R}$, can be arranged (cf.  
Eq.~\ref{eq:leffq2}) into a change of the normalization of the rate according to the replacement 
$V_{us}\to\tilde{V}_{us}=(1+\epsilon_L+\epsilon_R)V_{us}$, and of the axial  
coupling to the leptonic current by the factor $(1-2\epsilon_R)$. The former combination
involves a modification of $V_{us}$ which 
has been tightly constrained by testing CKM unitarity~\cite{Cirigliano:2009wk}. The latter
could be determined in SHD from the measured $g_1(0)\to\tilde{g}_1(0)=(1-2\epsilon_R)g_1(0)$
only if $g_1(0)$ was known accurately from QCD (for recent progress in the lattice 
see Refs.~\cite{Gockeler:2011ze,Green:2014vxa}).

On the other hand, the WC $\epsilon_{S,P,T}$ introduce new structures in the energy and angular
distributions. Restricting ourselves to $\mathcal{O}(v^2/\Lambda^2)$
(or linear in the WC), they appear from the interference of the NP terms with the SM and the contributions of the (pseudo)scalar and tensor operators
are suppressed by $m_\ell/\sqrt{q^2}$. Therefore, while the electronic channels can be analyzed
specifically to measure and study the normalization of the rates 
$|\tilde{V}_{us}\,f_1(0)|$ and the relevant form factors, the muonic modes could use
the information thus obtained to constrain the
(pseudo)scalar and tensor operators. Besides that, it is important to note 
that the pseudoscalar quark bilinear receives a kinematical $\mathcal{O}(q/M_1)$
suppression that largely neutralizes the sensitivity of SHD to $\epsilon_P$ (see however Ref.~\cite{Gonzalez-Alonso:2013ura}). 
For this reason, we center our discussion below on
the study of $\epsilon_S$ and $\epsilon_T$. 

We expand the contributions in the SM up to $\mathcal{O}(\delta)$,
but we keep only the leading terms in the NP terms. 
This implies a relative $\mathcal{O}(\delta^2)$ error in the SM predictions, which
we fix to a 5\% in all channels for definiteness, and an
uncertainty $\mathcal{O}(\delta)\sim10-20\%$ in the sensitivity
to NP that will not affect the conclusions of our analysis. 

\begin{table}
 \caption{\label{tab:RmueSMcomp} Comparison between the predictions of $R^{\mu e}$ in the SM 
 at NLO and experimental measurements for different SHD.} 

\begin{center}
    \begin{tabular}{ccccc}
      \hline 
      \hline
      &$\Lambda\to p$ & $\Sigma^{-}\to n$ & $\Xi^{0}\to \Sigma^{+}$ & $\Xi^{-}\to \Lambda$\\
      \hline 
      Expt. &$0.189(41)$&$0.442(39)$&$0.0092(14)$&$0.6(5)$  \\
      \hline 
      SM-NLO &$0.153(8)$&$0.444(22)$&$0.0084(4)$&$0.275(14)$\\
      \hline
      \hline
    \end{tabular}
  \end{center}
\end{table}

\paragraph{Bounds on scalar and tensor operators.-}
Let us now introduce the ratio:
\begin{equation}
R^{\mu e}=\frac{\Gamma(B_1\to B_2\,\mu^-\,\bar{\nu}_\mu)}{\Gamma(B_1\to B_2\,e^-\,\bar{\nu}_e)}. \label{eq:Rmue}
\end{equation} 
This observable is not only sensitive to lepton-universality violation
but also linearly sensitive to $\epsilon_S$ 
and $\epsilon_T$. In addition, one expects
the dependence on the form factors in the SM to simplify in the ratio. In fact, 
working at NLO we obtain:
\begin{eqnarray}
R^{\mu e}_{\rm SM} &=&\sqrt{1-\frac{m_\mu^2}{\Delta^2}}\left(1-\frac{9}{2}\,\frac{m_\mu^2}{\Delta^2}-4\frac{m_\mu^4}{\Delta^4}\right)\nonumber\\
&&+\frac{15}{2}\frac{m_\mu^4}{\Delta^4} 
{\rm arctanh}\left(\sqrt{1-\frac{m_\mu^2}{\Delta^2}}\right).\label{eq:RSM}
\end{eqnarray}
This is a remarkable result: up to a relative theoretical accuracy of 
$\mathcal{O}(\delta^2)$, $R^{\mu e}$ in the SM does not depend on any form factor. In Table~\ref{tab:RmueSMcomp} we compare the 
experimental ratios to the results predicted in the SM. 
As discussed above, the main reason for the large experimental errors is that most of the data in the muonic
channel is very old and scarce. At this level of precision, which generously covers
the theoretical accuracy attained by Eq.~(\ref{eq:RSM}), we observe that the experimental
data on $R^{\mu e}$ agrees with the SM.

One can now use this consistency of the data with the SM to set bounds on the WC
of the scalar and tensor operators, which generate the following non-standard contribution:
\begin{eqnarray}
R^{\mu e}_{\rm NP}\simeq \frac{\left(\epsilon_S\frac{f_S(0)}{f_1(0)}
+12\,\epsilon_T\,\frac{g_1(0)}{f_1(0)}\frac{f_T(0)}{f_1(0)}\right)}{(1-\frac{3}{2}\delta)\left(1+3\frac{g_1(0)^2}{f_1(0)^2}\right)}\,\Pi(\Delta,m_\mu), \label{eq:RmueNP}
\end{eqnarray}
where $\Pi(\Delta,m_\mu)$ is a phase-space integral:
\begin{eqnarray}
 &&\Pi(\Delta,m_\mu)=
 \frac{5}{2}\,\frac{m_\mu}{\Delta}\Bigg[\left(2+13\frac{m_\mu^2}{\Delta^2}\right)\sqrt{1-\frac{m_\mu^2}{\Delta^2}}\nonumber\\
 &&-3\left(4\,\frac{m_\mu^2}{\Delta^2}+\frac{m_\mu^4}{\Delta^4}\right)\,
 {\rm arctanh}\left(\sqrt{1-\frac{m_\mu^2}{\Delta^2}}\right)\Bigg]\,.
 \label{eq:PiNP}
\end{eqnarray}
It is particularly convenient to express the dependence on the WC in ``units'' of the
SM ratio:
\begin{eqnarray}
\frac{R^{\mu e}}{R^{\mu e}_{\rm SM}}=1+r_{S}\,\epsilon_S+r_T\,\epsilon_T, \label{eq:sensit}
\end{eqnarray}
where $r_{S,T}$ are dimensionless numbers encapsulating the net sensitivity to the WC.

\begin{table}[h]
  \caption{\label{tab:SHDdata}
  SHD data for $g_{1}(0)/f_{1}(0)$ and theoretical determinations of $f_{S,T}(0)/f_{1}(0)$
  at $\mu=2$ GeV used in this work. The corresponding $r_{S,T}$ are shown in the last two lines. 
  }
  \begin{center}
    \begin{tabular}{ccccc}
      \hline 
      \hline 
      &  $\Lambda\to p$ & $\Sigma^{-}\!\to n$ & $\Xi^{0}\!\to\Sigma^{+}$ & $\Xi^{-}\!\to\Lambda$\tabularnewline
      \hline 
      $g_{1}(0)/f_{1}(0)$ & $0.718(15)$ &$-0.340(17)$ & $1.210(50)$ & $0.250(50)$
      \tabularnewline
      $f_S(0)/f_1(0)$ & $1.90(10)$ & $2.80(14)$ & $1.36(7)$ &$2.25(11)$
      \tabularnewline
       $f_T(0)/f_1(0)$&$0.72$& $-0.28$&$1.22$&$0.22$
      \tabularnewline       
      \hline 
      $r_S$&$1.60$&$4.1$&$0.56$&$3.7$\\
      $r_T$ & $5.2$ & $1.7$& $7.2$&$1.1$\\
      \hline 
      \hline 
    \end{tabular}
  \end{center}
\end{table}

The values of the form factors that we use to calculate $r_{S,T}$ are given in Tab.~\ref{tab:SHDdata}. 
The ratio $g_1(0)/f_1(0)$ is measured from the angular distribution of the
electronic channels~\cite{Agashe:2014kda}. The scalar form factor can be obtained, 
up to electromagnetic corrections, using the conservation of vector current in QCD, 
$f_S(0)/f_1(0)=\Delta/(m_s-m_u)$~\cite{Gonzalez-Alonso:2013ura}. For the tensor
form factors we need to use model calculations~\cite{Ledwig:2010tu}, whose
errors are difficult to quantify. 
Nevertheless, it is interesting to note that the tensor form factor
for the neutron $\beta$-decay is predicted to be $1.22$, which is in the ballpark of
the values obtained in the lattice~\cite{Bhattacharya:2011qm,Green:2012ej,Bhattacharya:2013ehc,Green:2014vxa,Bali:2014nma}. 
This situation should be easily improved by future lattice calculations of the
hyperon decay tensor charges.

\begin{figure}[h]
\begin{center}
\includegraphics[scale=0.38]{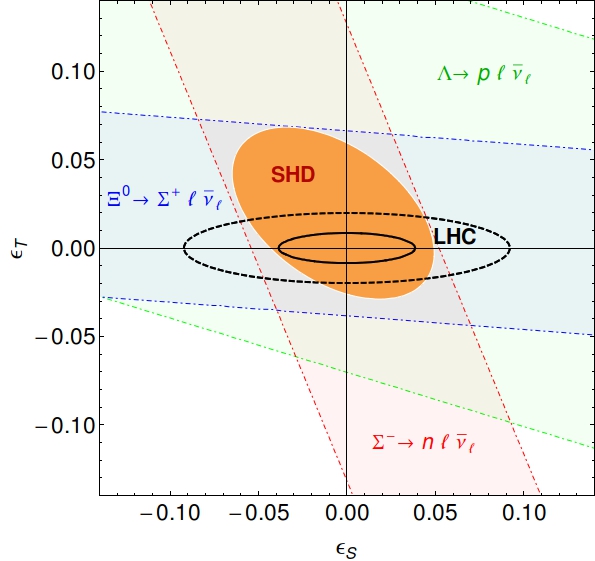}
\caption{\label{fig:CPST}
90\% CL constraints on $\epsilon_{S,T}$ at $\mu=2$ GeV from the measurements of $R^{\mu e}$ in different channels (dot-dashed lines)
and combined (filled ellipse). LHC bounds obtained from CMS data at $\sqrt{s}=$ 8 TeV (7 TeV) are represented by the black solid (dashed) ellipse.
\label{fig:C7p}}
\end{center}
\end{figure}

The sensitivities to $\epsilon_{S,T}$ exhibited by the SHD (last two lines of Tab.~\ref{tab:SHDdata})
are strongly channel-dependent. 
In Fig.~\ref{fig:CPST}, we show 90\% confidence level contours in the $(\epsilon_S,\,\epsilon_T)$ plane using a $\chi^2$
that includes the experimental measurements of $R^{\mu e}$ and where we propagate
the experimental and theoretical uncertainties of the SM predictions in quadratures.
For $r_{S,T}$ we use the values in Tab.~\ref{tab:SHDdata}. As we can see, 
even though the experimental data on $R^{\mu e}$ is not precise, the strong sensitivity
of SHD to NP leads to stringent bounds in $\epsilon_{S,T}$; namely:
\begin{equation}
\epsilon_S=0.003(40),\hspace{1cm}\epsilon_T=0.017(34)~,\label{eq:estbounds} 
\end{equation}
at 90\% C.L. Accounting for the running of $\epsilon_{S,T}$ on the renormalization scale $\mu$~\cite{Broadhurst:1994se},
and assuming natural values for the WC at $\mu=\Lambda$, these bounds translate into
$\Lambda\sim v\,(V_{us}\,\epsilon_{S,T})^{-1/2}\sim2-4$~TeV~\cite{Cirigliano:2012ab}.

\paragraph{Limits from LHC data.-}

As discussed above, the SMEFT allows to interpret model-independently high-
and low-energies searches of NP. In particular the cross-section $\sigma(pp\to e+{\rm MET}+X)$
with transverse mass higher than $\overline{m}_T$ is modified by non-standard $\bar{u}s\to e\bar{\nu}$
partonic interactions as follows:
\bea
\label{eq:sigmamt}
\sigma(m_T \!\!>\! \overline{m}_T) &=&
\sigma_W 
+ \sigma_S|\eS|^2 +  \sigma_T |\eT|^2 ~,%\nonumber 
\eea  
where $\sigma_W (\overline{m}_T)$ represents the SM contribution and $\sigma_{S,T}(\overline{m}_T)$ 
are new functions, which explicit form can be found in Ref.~\cite{Cirigliano:2012ab} 
(up to trivial flavor indexes changes).
Thus, comparing the observed events above $\overline{m}_T$ with the SM expectation
we can set bounds on $\epsilon_{S,T}$.
In particular, one (three) event is found with a transverse mass above $\overline{m}_T=1.5$~TeV
($1.2$~TeV) in the 20 fb$^{-1}$ (5 fb$^{-1}$) dataset recorded at $\sqrt{s}=$ 8 TeV
(7 TeV) by the CMS collaboration~\cite{Khachatryan:2014tva,Chatrchyan:2012meb}, in good
agreement with the SM background of $2.02\pm 0.26$ ($2.8\pm 1.0$) events. Using Eq.~\eqref{eq:sigmamt}
this agreement translates in the 90\% C.L. limits on $\epsilon_{S,T}$ shown in Fig.~\ref{fig:C7p}.
We use the MSTW2008 PDF sets evaluated at $Q^2=1$ TeV$^2$~\cite{Martin:2009iq} to calculate $\sigma_{S,T}$.
Further details can be found in Ref.~\cite{Cirigliano:2012ab}.

Fig.~\ref{fig:C7p} illustrates the interesting competition that future SHD measurements
could have with LHC searches of NP affecting $s\!\to\! u$ transitions. It is important to note that the dependence of the cross section~(\ref{eq:sigmamt})
on the WC is quadratic, whereas in SHD is linear. Besides reducing the 
sensitivity of the future collider searches of NP in this channel, one might also need to
consider possible cancellations with linear effects from dimension-8 operators in the SMEFT.

\paragraph{Conclusions and outlook.-} 
In summary, the most important features of SHD in relation to searches
of NP at TeV scales are: \textit{(i)}~The SHD are controlled by a small $SU(3)$-breaking
parameter, allowing for systematic expansions that lead to accurate expressions
in terms of a reduced dependence on form factors, cf. Eq.~(\ref{eq:RSM}).   
\textit{(ii)}~The interference of the (pseudo)scalar and tensor NP operators with the SM 
in the rate is chirally suppressed. Therefore, electronic modes
are well suited to measure normalization factors $|\tilde{V}_{us}\,f_1(0)|$,
NP-modified $\tilde{g}_1(0)$ and other form factors. \textit{(iii)}~The 
muonic modes, on the other hand, show a strong linear sensitivity to scalar and tensor contributions
that depend on the different combinations of form factors in each channel. This allows
to constrain them using SHD alone, with a precision that is competitive with the LHC data,
cf. Fig.~\ref{fig:CPST} and Eq.~(\ref{eq:estbounds}). 

Our hope is that the present study triggers a program of high-precision measurements 
of different observables in the SHD. Hyperons can be produced in great numbers
in current~\cite{PicciniNA62,Miwa:2012zz} and future facilities~\cite{Schonning:2014kza}. 
One may also wonder if better measurements could be extracted from the analysis of
the data collected in past experiments like HyperCP~\cite{White:2007zza}, KTeV and NA48. Any
development on the experimental side will directly improve the bounds on NP obtained 
in this work with an observable as simple as $R^{\mu e}$, and using data
with $\sim10-20\%$ relative errors.  

Future improvements on the experimental precision will need to be accompanied by 
similar efforts on the theory side. In particular, the inclusion of $\mathcal{O}(\delta^2)$ 
terms in the SM predictions would improve the accuracy to $\sim1\%-1\permil$. 
Besides that, further nonperturbative calculations of the tensor form factors would improve the assessment of the
sensitivity to $\epsilon_T$. Finally, it will be important
to perform this comprehensive analysis of the SHD in complementarity with the 
kaon decays. Work along these lines is in progress.

\paragraph{Acknowledgements.-}

We thank Sacha Davidson and Elizabeth Jenkins for useful discussions. The project was partially funded
by DOE grant DE-SC0009919 (H.-M.C.), the Lyon Institute of Origins, grant ANR-10-LABX-66 (M.G.A),
the People Programme (Marie Curie Actions) of the European Union's Seventh Framework Programme
(FP7/2007-2013) under REA grant agreement PIOF-GA-2012-330458 (J.M.C.) and the Spanish Ministerio
de Econom\'ia y Competitividad and European FEDER funds under the contract FIS2011-28853-C02-01 (J.M.C.).


\begin{thebibliography}{99}
%
%\cite{Weinberg:2009zz}
\bibitem{Weinberg:2009zz} 
  S.~Weinberg,
  %``V-A was the key,''
  J.\ Phys.\ Conf.\ Ser.\  {\bf 196}, 012002 (2009).
  %%CITATION = 00462,196,012002;%%
%  
%\cite{Cabibbo:1963yz}
\bibitem{Cabibbo:1963yz} 
  N.~Cabibbo,
  %``Unitary Symmetry and Leptonic Decays,''
  Phys.\ Rev.\ Lett.\  {\bf 10}, 531 (1963).
  %%CITATION = PRLTA,10,531;%%
%
%\cite{Cirigliano:2009wk}
\bibitem{Cirigliano:2009wk}
  V.~Cirigliano, J.~Jenkins and M.~Gonz\'alez-Alonso,
  %``Semileptonic decays of light quarks beyond the Standard Model,''
  Nucl.\ Phys.\ B {\bf 830} (2010) 95.
  %%CITATION = ARXIV:0908.1754;%%

%  
%\cite{Herczeg:2001vk}
\bibitem{Herczeg:2001vk} 
  P.~Herczeg,
  %``Beta decay beyond the standard model,''
  Prog.\ Part.\ Nucl.\ Phys.\  {\bf 46}, 413 (2001).
  %%CITATION = PPNPD,46,413;%%
%
%\cite{Severijns:2006dr}
\bibitem{Severijns:2006dr} 
  N.~Severijns, M.~Beck and O.~Naviliat-Cuncic,
  %``Tests of the standard electroweak model in beta decay,''
  Rev.\ Mod.\ Phys.\  {\bf 78}, 991 (2006).
  %%CITATION = NUCL-EX/0605029;%%  
%
%\cite{Bhattacharya:2011qm}
\bibitem{Bhattacharya:2011qm}
  T.~Bhattacharya, V.~Cirigliano, S.~D.~Cohen, A.~Filipuzzi, M.~Gonz\'alez-Alonso, M.~L.~Graesser, R.~Gupta and H.~-W.~Lin,
  %``Probing Novel Scalar and Tensor Interactions from (Ultra)Cold Neutrons to the LHC,''
  Phys.\ Rev.\ D {\bf 85} (2012) 054512.
  %%CITATION = ARXIV:1110.6448;%%
%  
%\cite{Cirigliano:2013xha}
\bibitem{Cirigliano:2013xha} 
  V.~Cirigliano, S.~Gardner and B.~Holstein,
  %``Beta Decays and Non-Standard Interactions in the LHC Era,''
  Prog.\ Part.\ Nucl.\ Phys.\  {\bf 71}, 93 (2013).
  %%CITATION = ARXIV:1303.6953;%%

%\cite{Gonzalez-Alonso:2013uqa}
\bibitem{Gonzalez-Alonso:2013uqa}
  O.~Naviliat-Cuncic and M.~González-Alonso,
  %``Prospects for precision measurements in nuclear $\beta$ decay at the LHC era,''
  Annalen Phys.\  {\bf 525} (2013) 600.
%  [arXiv:1304.1759 [hep-ph]].
  %%CITATION = ARXIV:1304.1759;%%
  %11 citations counted in INSPIRE as of 20 Dec 2014

%  
%\cite{Gonzalez-Alonso:2013ura}
\bibitem{Gonzalez-Alonso:2013ura} 
  M.~Gonz\'alez-Alonso and J.~Martin~Camalich,
  %``Isospin breaking in the nucleon mass and the sensitivity of $\beta$ decays to new physics,''
  Phys.\ Rev.\ Lett.\  {\bf 112}, no. 4, 042501 (2014).
  %%CITATION = ARXIV:1309.4434;%%
%
%\cite{Herczeg:1994ur}
\bibitem{Herczeg:1994ur} 
  P.~Herczeg,
  %``On the question of a tensor interaction in pi ---> e electron-neutrino gamma decay,''
  Phys.\ Rev.\ D {\bf 49}, 247 (1994).
  %%CITATION = PHRVA,D49,247;%%
%  
%\cite{Campbell:2003ir}
\bibitem{Campbell:2003ir} 
  B.~A.~Campbell and D.~W.~Maybury,
  %``Constraints on scalar couplings from pi+- ---> l+- nu(l),''
  Nucl.\ Phys.\ B {\bf 709}, 419 (2005).
  %%CITATION = HEP-PH/0303046;%%

%\cite{Cirigliano:2012ab}
\bibitem{Cirigliano:2012ab}
  V.~Cirigliano, M.~Gonzalez-Alonso and M.~L.~Graesser,
  %``Non-standard Charged Current Interactions: beta decays versus the LHC,''
  JHEP {\bf 1302} (2013) 046.
  %%CITATION = ARXIV:1210.4553;%%
  %21 citations counted in INSPIRE as of 20 Dec 2014

%\cite{Cirigliano:2011ny}
\bibitem{Cirigliano:2011ny} 
  V.~Cirigliano, G.~Ecker, H.~Neufeld, A.~Pich and J.~Portoles,
  %``Kaon Decays in the Standard Model,''
  Rev.\ Mod.\ Phys.\  {\bf 84}, 399 (2012).
  %%CITATION = ARXIV:1107.6001;%%
  
%\cite{Bernard:2006gy}
\bibitem{Bernard:2006gy} 
  V.~Bernard, M.~Oertel, E.~Passemar and J.~Stern,
  %``K(mu3)**L decay: A Stringent test of right-handed quark currents,''
  Phys.\ Lett.\ B {\bf 638}, 480 (2006).
  %%CITATION = HEP-PH/0603202;%%
  
%\cite{Bernard:2007cf}
\bibitem{Bernard:2007cf} 
  V.~Bernard, M.~Oertel, E.~Passemar and J.~Stern,
  %``Tests of non-standard electroweak couplings of right-handed quarks,''
  JHEP {\bf 0801}, 015 (2008).
  %%CITATION = ARXIV:0707.4194;%%
  
%\cite{Yushchenko:2004zs}
\bibitem{Yushchenko:2004zs} 
  O.~P.~Yushchenko, S.~A.~Akimenko, G.~I.~Britvich, K.~V.~Datsko, A.~P.~Filin, A.~V.~Inyakin, A.~S.~Konstantinov and V.~F.~Konstantinov {\it et al.},
  %``High statistic measurement of the K- ---> pi0 e- nu decay form-factors,''
  Phys.\ Lett.\ B {\bf 589}, 111 (2004).
  %%CITATION = HEP-EX/0404030;%%
  
    %\cite{Antonelli:2009ws}
\bibitem{Antonelli:2009ws} 
  M.~Antonelli, D.~M.~Asner, D.~A.~Bauer, T.~G.~Becher, M.~Beneke, A.~J.~Bevan, M.~Blanke and C.~Bloise {\it et al.},
  %``Flavor Physics in the Quark Sector,''
  Phys.\ Rept.\  {\bf 494}, 197 (2010).
  %%CITATION = ARXIV:0907.5386;%%
  
 %\cite{Carpentier:2010ue}
\bibitem{Carpentier:2010ue} 
  M.~Carpentier and S.~Davidson,
  %``Constraints on two-lepton, two quark operators,''
  Eur.\ Phys.\ J.\ C {\bf 70}, 1071 (2010).
  %%CITATION = ARXIV:1008.0280;%%
  

  %\cite{Antonelli:2008jg}
\bibitem{Antonelli:2008jg} 
  M.~Antonelli {\it et al.}  [FlaviaNet Working Group on Kaon Decays Collaboration],
  %``Precision tests of the Standard Model with leptonic and semileptonic kaon decays,''
  arXiv:0801.1817 [hep-ph].
  %%CITATION = ARXIV:0801.1817;%%  
  

  
%  
%\cite{Gaillard:1984ny}
\bibitem{Gaillard:1984ny} 
  J.~M.~Gaillard and G.~Sauvage,
  %``Hyperon Beta Decays,''
  Ann.\ Rev.\ Nucl.\ Part.\ Sci.\  {\bf 34}, 351 (1984).
  %%CITATION = ARNUA,34,351;%%
%
%\cite{Garcia:1985xz}
\bibitem{Garcia:1985xz} 
  A.~Garcia, P.~Kielanowski and A.~Bohm,
  %``The Beta Decay Of Hyperons,''
  Lect.\ Notes Phys.\  {\bf 222}, 1 (1985).
  %%CITATION = LNPHA,222,1;%%
    
%
%\cite{Cabibbo:2003cu}
\bibitem{Cabibbo:2003cu} 
  N.~Cabibbo, E.~C.~Swallow and R.~Winston,
  %``Semileptonic hyperon decays,''
  Ann.\ Rev.\ Nucl.\ Part.\ Sci.\  {\bf 53}, 39 (2003).
  %%CITATION = HEP-PH/0307298;%%

  %\cite{Kadeer:2005aq}
\bibitem{Kadeer:2005aq} 
  A.~Kadeer, J.~G.~Korner and U.~Moosbrugger,
  %``Helicity analysis of semileptonic hyperon decays including lepton mass effects,''
  Eur.\ Phys.\ J.\ C {\bf 59}, 27 (2009).
  %%CITATION = HEP-PH/0511019;%%  
 
%  
%\cite{Affolder:1999pe}
\bibitem{Affolder:1999pe} 
  A.~A.~Affolder {\it et al.}  [KTeV E832/E799 Collaboration],
  %``Observation of the decay Xi0 ---> Sigma+ e- anti-neutrino,''
  Phys.\ Rev.\ Lett.\  {\bf 82}, 3751 (1999).
  %%CITATION = PRLTA,82,3751;%%
  
%\cite{AlaviHarati:2001xk}
\bibitem{AlaviHarati:2001xk} 
  A.~Alavi-Harati {\it et al.}  [KTeV Collaboration],
  %``First measurement of form-factors of the decay Xi0 ---> Sigma+ e- anti-nu(e),''
  Phys.\ Rev.\ Lett.\  {\bf 87}, 132001 (2001).
  %%CITATION = HEP-EX/0105016;%%  
  
  %\cite{AlaviHarati:2005ev}
\bibitem{AlaviHarati:2005ev} 
  E.~Abouzaid {\it et al.}  [KTeV Collaboration],
  %``Observation of the decay Xi0 ---> Sigma+ mu- anti-nu(mu),''
  Phys.\ Rev.\ Lett.\  {\bf 95}, 081801 (2005).
  %%CITATION = HEP-EX/0504055;%%
  
  %\cite{Batley:2006fc}
\bibitem{Batley:2006fc} 
  J.~R.~Batley {\it et al.}  [NA48/I Collaboration],
  %``Measurement of the branching ratios of the decays Xi0 ---> Sigma+ e- anti-nu(e) and anti-Xi0 ---> anti-Sigma+ e+ nu(e),''
  Phys.\ Lett.\ B {\bf 645}, 36 (2007).
  %%CITATION = HEP-EX/0612043;%%
  
%\cite{Batley:2012mi}
\bibitem{Batley:2012mi} 
  J.~R.~Batley {\it et al.}  [NA48/1 Collaboration],
  %``Measurement of the branching ratio of the decay $\Xi^{0}\rightarrow \Sigma^{+} \mu^{-} \bar{\nu}_{\mu}$,''
  Phys.\ Lett.\ B {\bf 720}, 105 (2013).
  %%CITATION = ARXIV:1212.3131;%%  
 
%  
%\cite{Agashe:2014kda}
\bibitem{Agashe:2014kda} 
  K.~A.~Olive {\it et al.}  [Particle Data Group Collaboration],
  %``Review of Particle Physics (RPP),''
  Chin.\ Phys.\ C {\bf 38}, 090001 (2014).
  %%CITATION = CHPHD,C38,090001;%%
  
\bibitem{PicciniNA62}
M.~Piccini, talk at NA62 Physics Handbook Workshop, 2009.

%\cite{Schonning:2014kza}
\bibitem{Schonning:2014kza} 
  K.~Schönning {\it et al.}  [PANDA Collaboration],
  %``Antihyperon-hyperon production in antiproton-proton annihilations with PANDA,''
  J.\ Phys.\ Conf.\ Ser.\  {\bf 503}, 012013 (2014).
  %%CITATION = 00462,503,012013;%%

%\cite{Miwa:2012zz}
\bibitem{Miwa:2012zz} 
  K.~Miwa {\it et al.}  [J-PARC P40 Collaboration],
  %``Experimental plan of Sigma p scatterings at J-PARC,''
  EPJ Web Conf.\  {\bf 20}, 05001 (2012).
  %%CITATION = 00776,20,05001;%%

%\cite{Leung:1984ni}
\bibitem{Leung:1984ni}
  C.~N.~Leung, S.~T.~Love and S.~Rao,
  %``Low-Energy Manifestations of a New Interaction Scale: Operator Analysis,''
  Z.\ Phys.\ C {\bf 31} (1986) 433.
  %%CITATION = ZEPYA,C31,433;%%
  %222 citations counted in INSPIRE as of 12 Feb 2015
%
%\cite{Buchmuller:1985jz}
\bibitem{Buchmuller:1985jz}
  W.~Buchmuller and D.~Wyler,
  %``Effective Lagrangian Analysis of New Interactions and Flavor Conservation,''
  Nucl.\ Phys.\ B {\bf 268} (1986) 621.
  %%CITATION = NUPHA,B268,621;%%

  %\cite{Grzadkowski:2010es}
\bibitem{Grzadkowski:2010es}
  B.~Grzadkowski, M.~Iskrzynski, M.~Misiak and J.~Rosiek,
  %``Dimension-Six Terms in the Standard Model Lagrangian,''
  JHEP {\bf 1010} (2010) 085.
  %%CITATION = ARXIV:1008.4884;%%
  
%\cite{Jenkins:2013zja}
\bibitem{Jenkins:2013zja} 
  E.~E.~Jenkins, A.~V.~Manohar and M.~Trott,
  %``Renormalization Group Evolution of the Standard Model Dimension Six Operators I: Formalism and lambda Dependence,''
  JHEP {\bf 1310}, 087 (2013)
  [arXiv:1308.2627 [hep-ph]].
  %%CITATION = ARXIV:1308.2627;%%

%\cite{Jenkins:2013wua}
\bibitem{Jenkins:2013wua} 
  E.~E.~Jenkins, A.~V.~Manohar and M.~Trott,
  %``Renormalization Group Evolution of the Standard Model Dimension Six Operators II: Yukawa Dependence,''
  JHEP {\bf 1401}, 035 (2014).
  %%CITATION = ARXIV:1310.4838;%%  
  
%\cite{Alonso:2013hga}
\bibitem{Alonso:2013hga} 
  R.~Alonso, E.~E.~Jenkins, A.~V.~Manohar and M.~Trott,
  %``Renormalization Group Evolution of the Standard Model Dimension Six Operators III: Gauge Coupling Dependence and Phenomenology,''
  JHEP {\bf 1404}, 159 (2014)
  [arXiv:1312.2014 [hep-ph]].
  %%CITATION = ARXIV:1312.2014;%%
  
%\cite{Alonso:2014zka}
\bibitem{Alonso:2014zka} 
  R.~Alonso, H.~M.~Chang, E.~E.~Jenkins, A.~V.~Manohar and B.~Shotwell,
  %``Renormalization group evolution of dimension-six baryon number violating operators,''
  Phys.\ Lett.\ B {\bf 734}, 302 (2014)
  [arXiv:1405.0486 [hep-ph]].
  %%CITATION = ARXIV:1405.0486;%% 
  
  %\cite{Davidson:2010uu}
\bibitem{Davidson:2010uu} 
  S.~Davidson and S.~Descotes-Genon,
  %``Minimal Flavour Violation for Leptoquarks,''
  JHEP {\bf 1011}, 073 (2010)
  [arXiv:1009.1998 [hep-ph]].
  %%CITATION = ARXIV:1009.1998;%%

%\cite{Weinberg:1958ut}
\bibitem{Weinberg:1958ut}
  S.~Weinberg,
  %``Charge symmetry of weak interactions,''
  Phys.\ Rev.\  {\bf 112} (1958) 1375.
  %%CITATION = PHRVA,112,1375;%%

%\cite{Georgi:1982jb}
\bibitem{Georgi:1982jb} 
  H.~Georgi,
  %``Lie Algebras In Particle Physics. From Isospin To Unified Theories,''
  Front.\ Phys.\  {\bf 54}, 1 (1982).
  %%CITATION = FRPHA,54,1;%%
  
  %\cite{Krause:1990xc}
\bibitem{Krause:1990xc} 
  A.~Krause,
  %``Baryon Matrix Elements of the Vector Current in Chiral Perturbation Theory,''
  Helv.\ Phys.\ Acta {\bf 63}, 3 (1990).
  %%CITATION = HPACA,63,3;%% 

%\cite{Jenkins:1990jv}
\bibitem{Jenkins:1990jv} 
  E.~E.~Jenkins and A.~V.~Manohar,
  %``Baryon chiral perturbation theory using a heavy fermion Lagrangian,''
  Phys.\ Lett.\ B {\bf 255}, 558 (1991).
  %%CITATION = PHLTA,B255,558;%%
  
%\cite{Geng:2008mf}
\bibitem{Geng:2008mf} 
  L.~S.~Geng, J.~Martin Camalich, L.~Alvarez-Ruso and M.~J.~Vicente Vacas,
  %``Leading SU(3)-breaking corrections to the baryon magnetic moments in Chiral Perturbation Theory,''
  Phys.\ Rev.\ Lett.\  {\bf 101}, 222002 (2008).
  %%CITATION = ARXIV:0805.1419;%%  

%\cite{Ledwig:2014rfa}
\bibitem{Ledwig:2014rfa} 
  T.~Ledwig, J.~Martin Camalich, L.~S.~Geng and M.~J.~V.~Vacas,
  %``Octet-baryon axial-vector charges and SU(3)-breaking effects in the semileptonic hyperon decays,''
  Phys.\ Rev.\ D {\bf 90}, 054502 (2014).
  %%CITATION = ARXIV:1405.5456;%%  

 %\cite{Geng:2013xn}
\bibitem{Geng:2013xn} 
  L.~Geng,
  %``Recent developments in SU(3) covariant baryon chiral perturbation theory,''
  Front.\ Phys.\ China {\bf 8}, 328 (2013).
  %%CITATION = ARXIV:1301.6815;%%   
 
%\cite{Guadagnoli:2006gj}
\bibitem{Guadagnoli:2006gj}
  D.~Guadagnoli, V.~Lubicz, M.~Papinutto and S.~Simula,
  %``First Lattice QCD Study of the Sigma ---> n Axial and Vector Form Factors with SU(3) Breaking Corrections,''
  Nucl.\ Phys.\ B {\bf 761}, 63 (2007).  %%CITATION = HEP-PH/0606181;%%

%\cite{Sasaki:2008ha}
\bibitem{Sasaki:2008ha}
  S.~Sasaki and T.~Yamazaki,
  %``Lattice study of flavor SU(3) breaking in hyperon beta decay,''
   Phys.\ Rev.\ D {\bf 79}, 074508 (2009).  
   %%CITATION = ARXIV:0811.1406;%%
 
%\cite{Sasaki:2012ne}
\bibitem{Sasaki:2012ne}
  S.~Sasaki,
  %``Hyperon vector form factor from 2+1 flavor lattice QCD,'' 
  Phys.\ Rev.\ D {\bf 86}, 114502 (2012).
  %%CITATION = ARXIV:1209.6115;%%  
  
%\cite{Ecker:1989yg}
\bibitem{Ecker:1989yg} 
  G.~Ecker, J.~Gasser, H.~Leutwyler, A.~Pich and E.~de Rafael,
  %``Chiral Lagrangians for Massive Spin 1 Fields,''
  Phys.\ Lett.\ B {\bf 223}, 425 (1989).
  %%CITATION = PHLTA,B223,425;%%
  
  %\cite{Masjuan:2012sk}
\bibitem{Masjuan:2012sk} 
  P.~Masjuan, E.~Ruiz Arriola and W.~Broniowski,
  %``Meson dominance of hadron form factors and large-Nc phenomenology,''
  Phys.\ Rev.\ D {\bf 87}, 014005 (2013).
  %%CITATION = ARXIV:1210.0760;%%
  
%\cite{FloresMendieta:1996np}
\bibitem{FloresMendieta:1996np} 
  R.~Flores-Mendieta, A.~Garcia, A.~Martinez and J.~J.~Torres,
  %``Radiative corrections to the semileptonic Dalitz plot with angular correlation between polarized decaying and emitted hyperons,''
  Phys.\ Rev.\ D {\bf 55}, 5702 (1997).
  %%CITATION = PHRVA,D55,5702;%% 
  
%\cite{Gockeler:2011ze}
\bibitem{Gockeler:2011ze} 
  M.~Gockeler {\it et al.}  [QCDSF/UKQCD Collaboration],
  %``Baryon Axial Charges and Momentum Fractions with N_f=2+1 Dynamical Fermions,''
  PoS LATTICE {\bf 2010}, 163 (2010).
  %%CITATION = ARXIV:1102.3407;%%  
  
%\cite{Green:2014vxa}
\bibitem{Green:2014vxa} 
  J.~Green,
  %``Hadron Structure from Lattice QCD,''
  arXiv:1412.4637 [hep-lat].
  %%CITATION = ARXIV:1412.4637;%%   
 
%\cite{Ledwig:2010tu}
\bibitem{Ledwig:2010tu} 
  T.~Ledwig, A.~Silva and H.~C.~Kim,
  %``Tensor charges and form factors of SU(3) baryons in the self-consistent SU(3) chiral quark-soliton model,''
  Phys.\ Rev.\ D {\bf 82}, 034022 (2010).
  %%CITATION = ARXIV:1004.3612;%%
 
   
  %\cite{Green:2012ej}
\bibitem{Green:2012ej} 
  J.~R.~Green, J.~W.~Negele, A.~V.~Pochinsky, S.~N.~Syritsyn, M.~Engelhardt and S.~Krieg,
  %``Nucleon Scalar and Tensor Charges from Lattice QCD with Light Wilson Quarks,''
  Phys.\ Rev.\ D {\bf 86}, 114509 (2012).
  %%CITATION = ARXIV:1206.4527;%%
  
  %\cite{Bhattacharya:2013ehc}
\bibitem{Bhattacharya:2013ehc} 
  T.~Bhattacharya, S.~D.~Cohen, R.~Gupta, A.~Joseph, H.~W.~Lin and B.~Yoon,
  %``Nucleon Charges and Electromagnetic Form Factors from 2+1+1-Flavor Lattice QCD,''
  Phys.\ Rev.\ D {\bf 89}, 094502 (2014).
  %%CITATION = ARXIV:1306.5435;%%
  
  %\cite{Bali:2014nma}
\bibitem{Bali:2014nma} 
  G.~S.~Bali, S.~Collins, B.~Glässle, M.~Göckeler, J.~Najjar, R.~H.~Rödl, A.~Schäfer and R.~W.~Schiel {\it et al.},
  %``Nucleon isovector couplings from $N_f=2$ lattice QCD,''
  arXiv:1412.7336 [hep-lat].
 
%\cite{Broadhurst:1994se}
\bibitem{Broadhurst:1994se}
  D.~J.~Broadhurst and A.~G.~Grozin,
  %``Matching QCD and HQET heavy - light currents at two loops and beyond,''
  Phys.\ Rev.\ D {\bf 52} (1995) 4082.
  %%CITATION = HEP-PH/9410240;%%
  
%\cite{Khachatryan:2014tva}
\bibitem{Khachatryan:2014tva}
  V.~Khachatryan {\it et al.}  [CMS Collaboration],
  %``Search for physics beyond the standard model in final states with a lepton and missing transverse energy in proton-proton collisions at $\sqrt{s}$ = 8 TeV,''
  arXiv:1408.2745 [hep-ex].
  %%CITATION = ARXIV:1408.2745;%%
  %17 citations counted in INSPIRE as of 12 Feb 2015  

%\cite{Chatrchyan:2012meb}
\bibitem{Chatrchyan:2012meb}
  S.~Chatrchyan {\it et al.}  [CMS Collaboration],
  %``Search for leptonic decays of $W$ ' bosons in $pp$ collisions at $\sqrt{s}=7$ TeV,''
  JHEP {\bf 1208} (2012) 023
  [arXiv:1204.4764 [hep-ex]].
  %%CITATION = ARXIV:1204.4764;%%
  %71 citations counted in INSPIRE as of 20 Feb 2015

%\cite{Martin:2009iq}
\bibitem{Martin:2009iq}
  A.~D.~Martin, W.~J.~Stirling, R.~S.~Thorne and G.~Watt,
  %``Parton distributions for the LHC,''
  Eur.\ Phys.\ J.\ C {\bf 63} (2009) 189
  [arXiv:0901.0002 [hep-ph]].
  %%CITATION = ARXIV:0901.0002;%%
  %2553 citations counted in INSPIRE as of 20 Feb 2015
  
  %\cite{White:2007zza}
\bibitem{White:2007zza} 
  C.~White [HyperCP Collaboration],
  %``Results from the HyperCP experiment (FNAL E871),''
  AIP Conf.\ Proc.\  {\bf 917}, 330 (2007).
  %%CITATION = APCPC,917,330;%%


\end{thebibliography}
\end{document}